\documentclass{appolb}
\usepackage{epsfig,amssymb,xspace,color}
\newcommand{\ave}[1]{\left\langle #1 \right\rangle}
\newcommand{\sqrts}{\sqrt{s}}
  \newcommand{\lqcd}{\Lambda_{QCD}}

\newcommand{\nc}{\newcommand}
\nc{\req}[1]{Eq.\,(\ref{#1})}  \nc{\eps}{\varepsilon}
\nc{\beq}{\begin{equation}}     \nc{\beql}[1]{\begin{equation}\label{#1}}
\nc{\eeq}{\end{equation}}       \nc{\rf}[1]{figure  \ref{#1}}
\nc{\beqa}{\begin{eqnarray}}   \nc{\eeqa}{\end{eqnarray}}
     \nc{\pathlaptop}{/home/rafelski/figure/}
     \nc{\pathletes}{/users/lpthe/jletes/bookraf/figures/}
     \nc{\pathnow}{}
 \def\lessim{\lower.5ex\hbox{$\; \buildrel < \over \sim \;$}}
\def\gtrsim{\lower.5ex\hbox{$\; \buildrel > \over \sim \;$}}

\usepackage{graphicx}

\begin{document}
\title{Strange quark matter:Business as usual or phase transition?
\thanks{Theory summary talk at SQM2011, Krakow}}%
\author{Giorgio Torrieri
\address{FIAS,
  J.W. Goethe Universit\"at, Frankfurt am Main, Germany}}
\maketitle
\begin{abstract}
We give an overview of some results presented at the Strange Quark Matter 2011 conference in Krakow, and interpret them in light of the search for the deconfinement QCD phase transition in heavy ion collisions
\end{abstract}
  
\section{Introduction: Why are we here}
My fellow summary speaker \cite{safarik} has asked what the \underline{s} in \underline{s}QM stands for.      The historic meaning of it is of course strangeness, but in this conference we had many excellent talks on topics which have very little to do with strange quarks: Jets, flow of light particles, quarkonium, percolation, and so on.   In fact, since the spread of talks presented in this conference matches the spread of talks of the larger ``Quark Matter'' series of conferences, another possible answer is that sQM stands for ``small Quark matter'' \cite{safarik}.

Is it true, therefore, that one can acceptably simulate an sQM proceeding by taking a QM proceeding and randomly accepting a talk every three or four?   I believe that a careful listener not just of the talks, but of the discussions, questions and so on will notice another difference:  QM is what it says it is... the study of quark matter in all its manifestations.   Historically, the sQM conference series has started with a more specific goal: Finding the deconfinement phase transition through changes in the chemistry of the system produced in heavy ion collisions.

Originally, the only observable where this was applicable was strangeness enhancement.   Other observables, however, have been deemed interesting in this regard as well.   Therefore, while the {\em content} of the conference has broadened, I believe the {\em focus} has largely remained.   This summary is written with this in mind, and hence looks at each experimental result and theoretical argument from the point of view of ``how far does it advance us to the goal of observing a deconfinement transition?''

In general, how does one look for a phase transition?  Since this is a fundamentally thermodynamic concept, one can look for inspiration in other areas of statistical mechanics: First, one needs a ``large'' system, as close as possible to the thermodynamic limit, where the total volume becomes a linearly scaling ``normalization factor'', and decouples from any ``intensive'' properties of the system (temperature, density of entropy, density of conserved charges, and so on).  Then, one looks for a scaling violation in an observable which jumps when one of these intensive quantities, or their derivatives, experience a discontinuity.

\begin{figure*}[tb]
\centerline{\epsfig{file=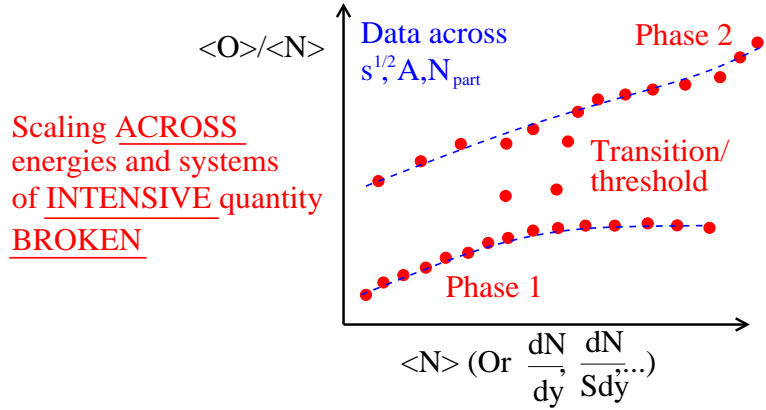,width=10cm}}
\caption{\label{goodsig} (Color online) 
What we hope to find
}
\end{figure*}

Our focusing on mostly, but not entirely, on heavy ion collisions is our way of reaching the thermodynamic limit in the laboratory.   Even before the ``perfect fluid'' discovery, it was always clear that in a heavy ion collision, the Knudsen number $l_{mfp}/L \sim \eta/(sTL) \leq 1$ and hence {\em some kind} of local equilibrium was likely to be achieved.   As for the scaling violation, Fig. \ref{goodsig} illustrates how it would work in practice: A graph where the ``x axis'' is some bulk observable of the system ($dN/dy,1/SdN/dy$ and so on), while the ``y axis'' is the observable of interest.  When all energies and system sizes are put on this graph (note that one should be able to meaningfully compare different energies and system sizes), a clear scaling violation emerges, indicated a change in {\em intensive} degrees of freedom.
I believe finding and defining such a signature is one of our main goals, if not our main goal.  I think we are not there yet, so in a sense the answer given to the question in the title is ``business as usual'', {\em but} there are interesting hints that a graph like Fig. \ref{goodsig} is in sight.  I remain optimistic that a graph such as Fig. \ref{goodsig} is obtainable, and I believe that significant strides in the direction of obtaining it were made in this meeting.   The next few sections will examine the talks presented, always with Fig. \ref{goodsig} in mind.    
\section{What are we looking for?  The QCD phase diagram}
Of course, the first step in our journey involves specifying the objective we are looking for, i.e. defining the QCD phase diagram.   Naively, at $T\ll T_c$ we should expect a weakly interacting gas of hadrons, either ``light'' (w.r.t. $\lqcd$) pions or heavy mesons and baryons.   At $T \sim T_c$ we have a  transition to partonic degrees of freedom. 
At $T \gg T_c$, we should have a weakly interacting gas of quarks and gluons.

The devil is, of course, in the details:  Lattice results \cite{gavai,schmidt} seem to show that, at low chemical potential one has a cross-over rather than a phase transition:  While intensive quantities jump (keeping our hope for Fig. \ref{goodsig}), their jump is continuous and differentiable throughout even for volumes approaching infinity.
It is however widely expected that the transition becomes first order at high chemical potential, allowing for the experimentally ``spectacular'' signature of a critical point and associated critical behavior.
\begin{figure}[t]
\begin{center}
\epsfig{width=12cm,figure=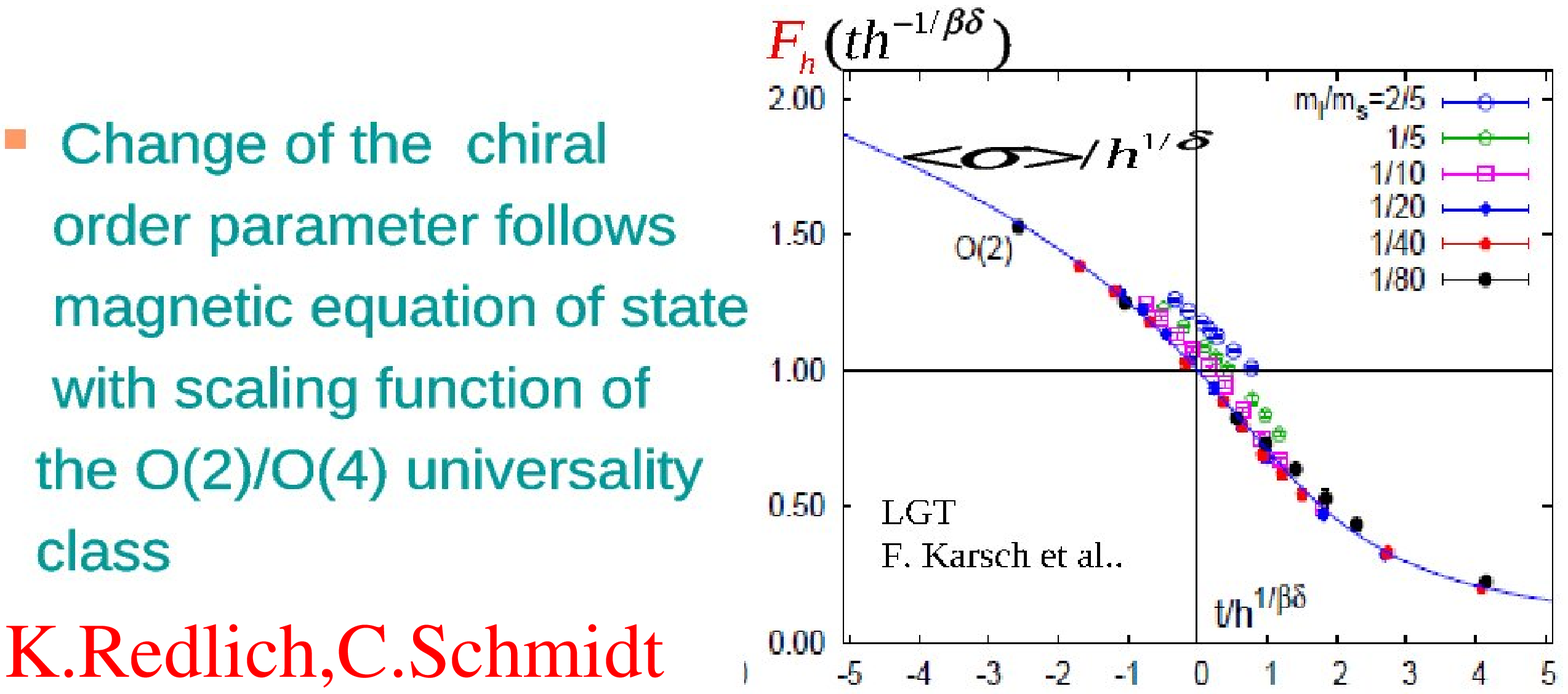}
\caption{(color online) numerical evidence for the critical point from universality-based arguments and lattice data \label{redlich} }
\end{center}
\end{figure}

Unfortunately, this belief can not as yet be rigorously checked theoretically, since \cite{gavai,schmidt} at high chemical potentials lattice calculations fail to converge because of the sign problem, yet the system is still strongly coupled and hence not amenable to perturbative techniques.
The belief in the critical point comes from two classes of evidence:  The wide variety of QCD effective theories exhibiting critical points \cite{costa}, and lattice-based evidence that QCD with 2+1 flavors is in the $O(4)$ universality class \cite{redlich,schmidt} (Fig. \ref{redlich}), which exhibits a pseudo-critical point.
\begin{figure}[t]
\begin{center}
\epsfig{width=13cm,figure=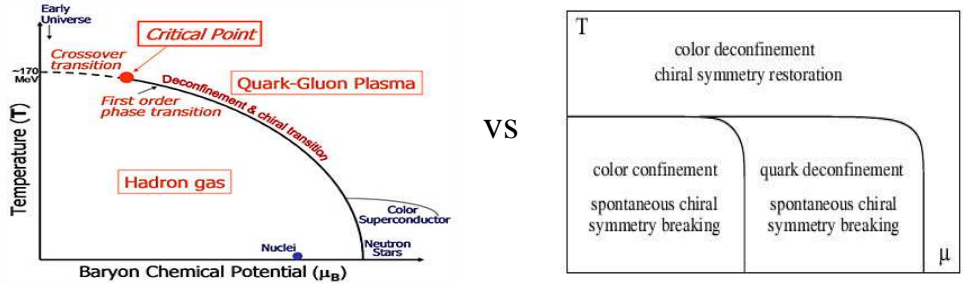}
\caption{(color online) A critical point or new phases? \label{satz}}
\end{center}
\end{figure}
Since the numerical evidence is not conclusive, and effective theories are very far from capturing the conceptual subtlety of QCD, the belief in the critical point remains a belief at this point, and surprises can not be excluded.
For example, \cite{satz} (Fig. \ref{satz} right panel) has hypothesized that instead of a critical point the system exhibits a {\em triple point}, with a new chirally broken phase of deconfined or semideconfined (percolating) constituent quarks.

Theoretically, such ideas are very interesting, but, unlike a critical point, we are currently lacking a {\em phenomenology} which would enable us to look for these phases in heavy ion experimental data.   Experiments such as FAIR,NICA,SHINE and RHIC low energy \cite{bleicher,sorin,gazdzicki,odyniec} might therefore be going into theoretically uncharted waters.

This is a good point to remind ourselves that there is more to heavy ion phenomenology than heavy ion collisions.
The ultimate acts of strangeness enhancement, in fact, might be in the sky around us:  \cite{sagert}'s effort to incorporate strangeness into the physics of supernovae and proto-neutron stars reminds us that our existence (as organic beings many of whose elements were created in supernova explosions) could literally owe something to strange quark matter.
Alternatively, we might all die thanks to strange quarks \cite{labun}, as a blob of cold strange matter from outer space will destroy our planet.
Last but not least, we will get a spectacular signature of cosmic strange quark matter in the form of exotica in particle accelerators \cite{steinheimer}.

Just as heavy ion physicists should remember that evidence of phase transitions need not come from accelerator-based experiments, astrophysicists interested in QCD should remember they might be looking for phase transitions too.  And that we {\em do not} conclusively know what lurks at the high chemical potential end of the phase diagram, the region most of interest to astrophysicists.  There might be surprises relevant to them.
\section{Global characteristics of the system \label{secglobal}}
The discussion around Fig. \ref{goodsig} makes it apparent that before looking for ``interesting probes'' one needs a comparison standard, the x-axis of Fig. \ref{goodsig}.

\begin{figure}[t]
\epsfig{file=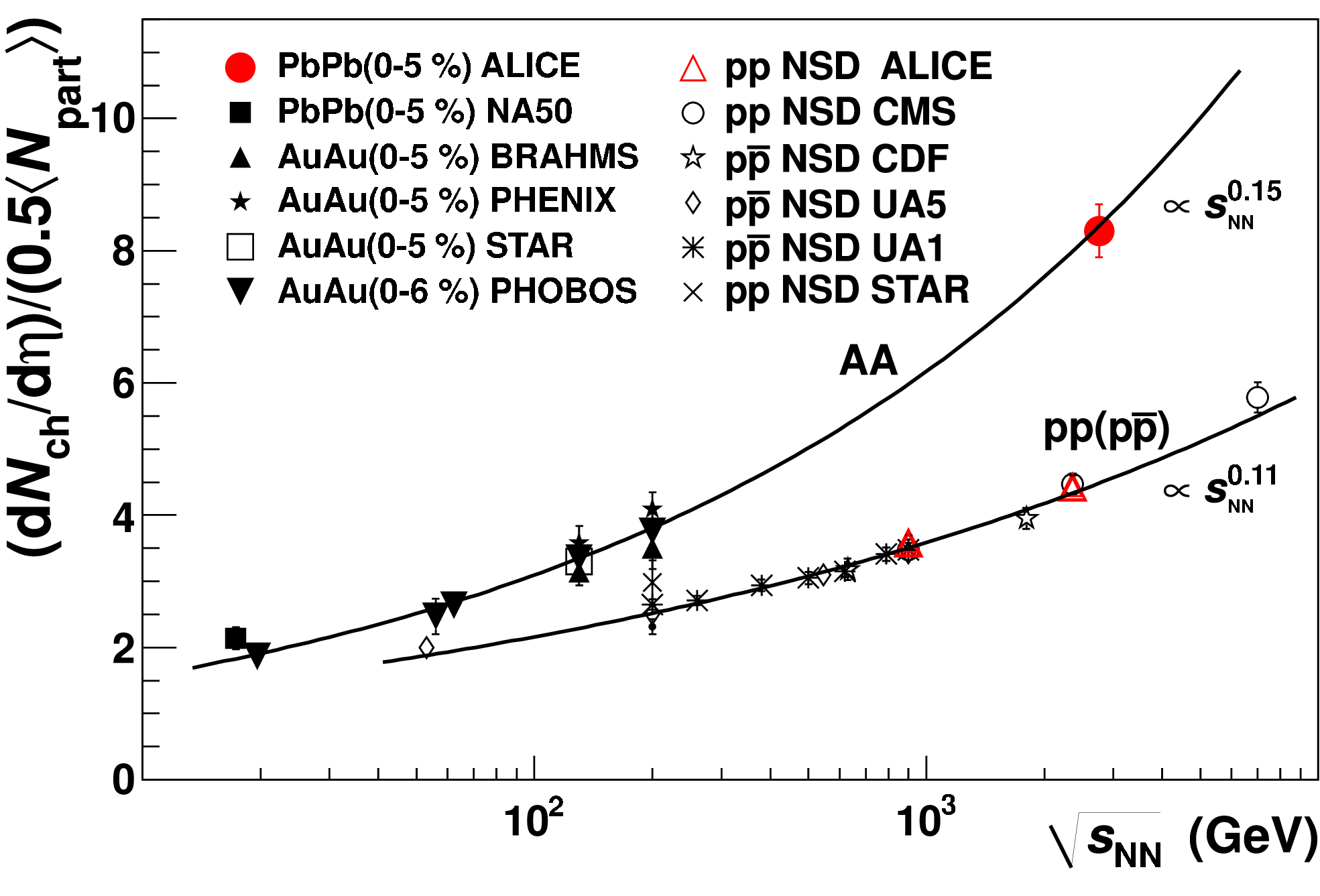,width=6cm}
\epsfig{file=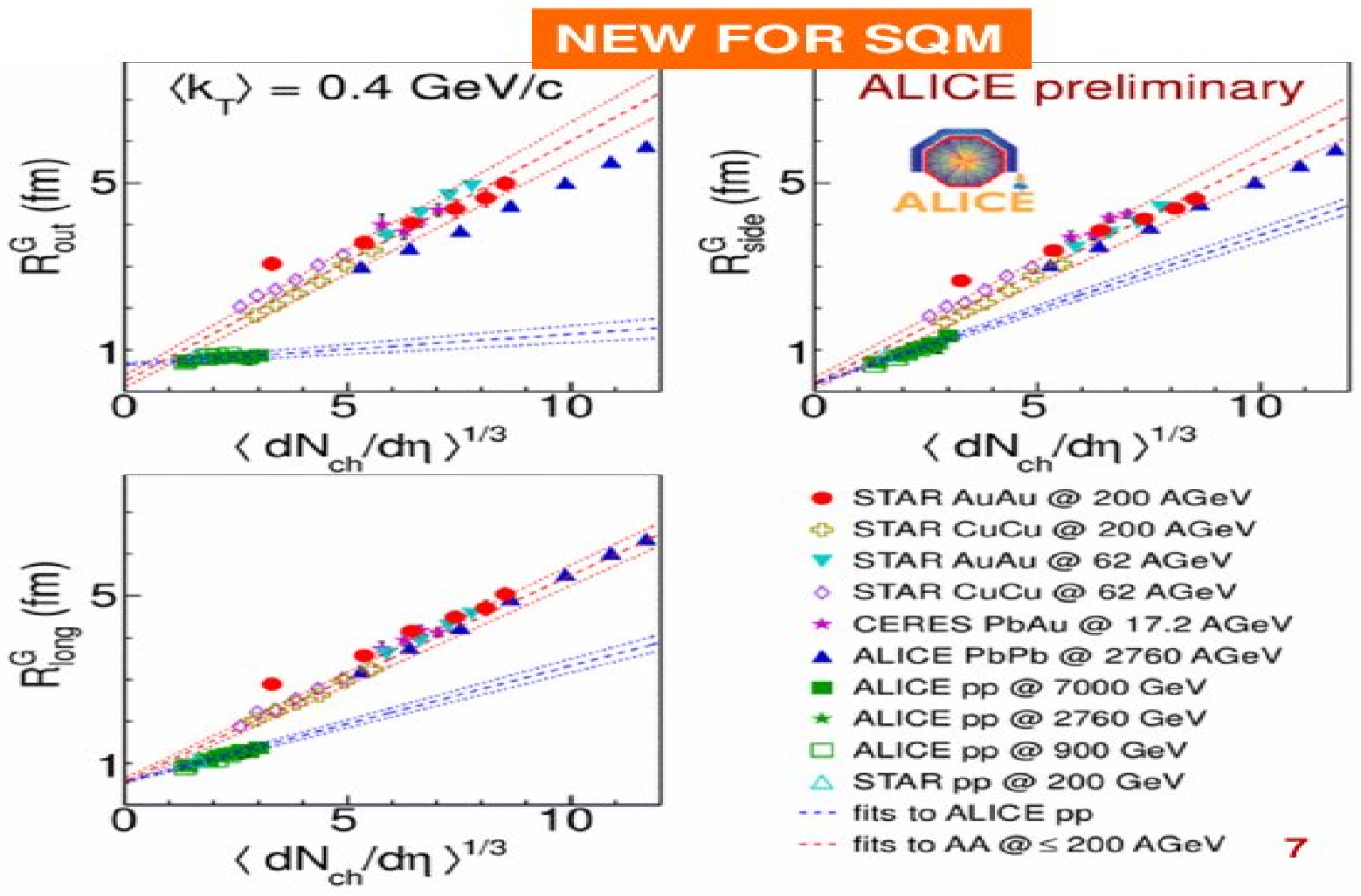,width=6cm}
\caption{\label{alicemult} (left panel) event multiplicity at mid-rapidity plotted against energy (right panel) HBT radii in $p-p$ and $A-A$ collisions (ALICE collaboration) }
\end{figure}

As Fig. \ref{alicemult} (left panel) shows \cite{prino,petrov} shows, this is a rather subtle issue, as the LHC has {\em conclusively} established that multiplicity per participant grows with energy as a power-law.   This increase, however, is only apparent at the LHC.   In other words, when comparing between different energies (e.g., when measuring strangeness enhancement \cite{prino}), one has to distinguish between ``the interesting observable changes across these energies'' from ``the system as a whole, including the interesting observable changes across these energies''.
ALICE \cite{prino} has also conclusively convinced us that soft physics at $p-p$ is different from $A-A$ even when the multiplicity of the event is the same (Fig. \ref{alicemult} right panel).  While both $p-p$ and $A-A$ source radii $\sim (dN/dy)^{1/3}$, the scaling constant is very different, so events with the comparable multiplicity in $p-p$ and $A-A$ have very different HBT radii.
Events of {\em different systems in } $A-A$, on the other hand, scale very well, so semicentral Pb-Pb collisions at the LHC have comparable HBT radii to central collisions at RHIC.  p-p collisions, accordingly, seem to scale well too.

This is a very nice result, whose theoretical interpretation is still uncertain.  As shown by \cite{bozek}, while hydrodynamics is still not {\em completely} reproducing HBT radii, it does a reasonable job at LHC A-A, does a worse job at RHIC A-A, and fails $p-p$.  This {\em certainly} means that A-A is, intensively, a very different system than p-p even when globally these two events have the same multiplicity.  It probably means that we still have not fully understood freeze-out, and its relation to deconfinement (Is a better fit at the LHC \cite{bozek} due to the fact that the initial temperature there is {\em well} above the deconfined region?).
Another crucial question which sometimes gets ignored when interpreting the newly released data on multiplicity is whether there really is proportionally ``more soft stuff'' (which thermalizes and is part of the collective evolution), or is there more ``unthermalized jetty corona'' contributing more soft particles (so, perhaps, the really soft stuff still follows the $dN/dy \sim \ln \sqrts$ scaling).   The answer is {\em not} trivial (remember that jet production changes strongly with $\sqrts$, and soft particle production seems to be very far from both the Landau and the Bjorken limits), but it {\em seems} that soft observables \cite{prino,bozek,huovinen} scale with $dN/dy$ rather than $\sqrts$, indicating the first alternative is perhaps closer to the truth (more data, such as scaling of $\ave{p_T}$ or limiting fragmentation, is however needed to confirm this).

More generally, disentangling violations of scaling of intensive quantities from modifications of total system size, chemical potential at mid-rapidity, transparency and rapidity intervals becomes a challenge when distinguishing ``high energy'' (top RHIC, LHC) from lower energy regimes.  Any ``signatures of deconfinement'', such as those hinted by \cite{gazdzicki,odyniec,kasia} in their respective talks, must be examined in this prospective, a work still in progress.
\section{Thermal models and their uses}
\subsection{Applicability}
The fact that statistical models fits heavy ion data is certainly not something this conference encounters for the first time.
The interpretation, and scope of validity, are however controversial, again not a new controversy at meetings like this.

First of all, there \cite{pbm} are those who believe that statistical models only work for large systems, and chemical equilibrium is most likely connected to the onset of a strongly interacting phase accompanying a phase transition.
On the other hand, others \cite{oeschler,becattini,biro} would extend the applicability of statistical models to much smaller systems, and interpret their validity as reflecting something fundamental about QCD in general.

Giving a straight-forward answer to this is not possible at the moment.  While strangeness enhancement and HBT show $p-p$ and $A-A$ are {\em very different} at all energies \cite{prino}, some things look more thermalized in $A-A$ (charmed quarks \cite{pbm},multistrange hadrons \cite{prino}) but others (resonances, baryons at ALICE \cite{prino}) might look more thermalized in $p-p$ \cite{blume}.

We might therefore need to be creative with defining observables sensitive to thermalization.  Here, several talks proposed observables sensitive to near-equilibrium behavior in small systems, such as heavy quark energy loss \cite{vogel} and hydrodynamic-type behavior ($v_n$,ridges) in $p-p$ collisions \cite{bozek,bialkowska}, and of course fluctuations and higher cumulants \cite{gavai,redlich,ritter}.   It will be interesting to see to what extend these observables can be measured for a range of energies and system sizes.
\begin{figure}[t]
\epsfig{file=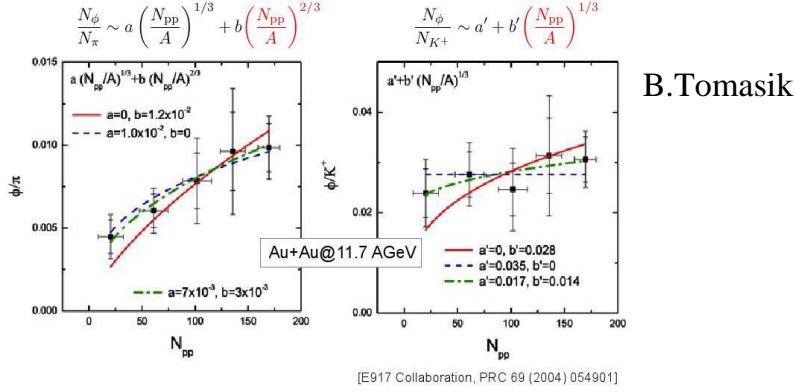,width=11cm}
\caption{\label{tomasik} Evidence of lack of equilibration in $\phi/\pi$ within a context of catalytic reactions, as inferred from $N_{part}$ scaling }
\end{figure}
If, however, thermalization certainly needs to be experimentally constrained better, {\em lack of thermalization} also needs to be demonstrated (beyond stating ``in my favorite model says observable X is not thermalized!''), precisely because the range of applicability of statistical models is so uncertain.  \cite{tomasik} and Fig. \ref{tomasik} show a very nice way to do this, using the fact that it is only in an equilibrated system that correlation volume is the same for {\em all} particle abundances.   \cite{tomasik} discussed the $\phi$ abundance, but other particles might be amenable to this treatment.  

Other than for its intrinsic value, the question of thermalization in smaller systems might be useful in our search for the critical point:  If smaller systems get thermalized, their freezeout temperature seems to be {\em higher} than the corresponding freezeout temperature of larger systems \cite{gazdzicki}.  Physically, this is natural if freezeout happens at a critical Knudsen number rather than a critical temperature.

\begin{figure}[t]
\begin{center}
\epsfig{width=10cm,figure=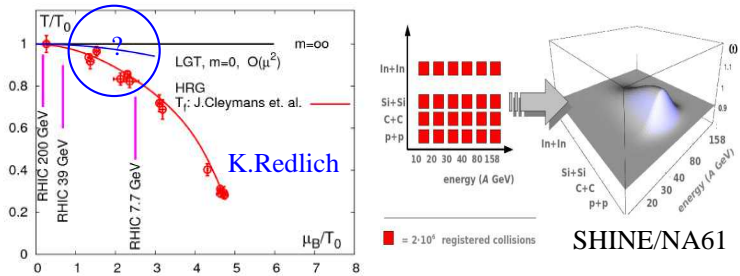}
\caption{(color online) Left-panel: An ``experimental'' $T-\mu$ graph superimposed with the expectation of what the phase diagram looks like, illustrating the possibility we might miss a critical point \cite{redlich,mall}.  Scans in system size \cite{gazdzicki} might provide a solution  \label{redlichcrit} }
\end{center}
\end{figure}

{\em If} the deviation from equilibrium is {\em not} consequently larger, \cite{mall,gazdzicki} proposes to use smaller systems to widen the available  $T-\mu$ area of search for the critical point, thereby avoiding the apparent mismatch between the freezeout curve and the QCD phase diagram pointed out in \cite{mall,redlich} (Fig. \ref{redlichcrit}).
Disentangling deviation from equilibrium from approach to critical point in a model-independent way will, however, be a challenge.

A discussion of the applicability and interpretation of the statistical model must also include a discussion of Tsallis statistics \cite{biro,urmossy}. The inclusion of the Tsallis parameter $q$ allows to confirm the extension of the statistical description to systems as small as $e^+ e^-$, and also describe high $p_T$ particle production which is commonly assumed to be controlled by pQCD and fragmentation.

A cynic would say that, since half the world scales exponentially and the other half scales as a power law, it is not surprising that a distribution such as the Tsallis one describes the world!  To defeat such cynicism, one would have to find a good physical interpretation for the Tsallis parameter $q$, preferably one with predictive power (how it varies with energy and system size).
Thus, \cite{biro,urmossy} would interpret the Tsallis parameter as correlations from ``apparent thermalization'' due to classical fields, while \cite{biro,cleymans} interpret it as ``meta-statistics'' due to an ensemble of thermalization temperatures and volumes (which in turn might give a high $p_T$ tail \cite{gazdzicki}).

The phenomenological way forward for these discussions might have to be fluctuations and higher cumulants (see section \ref{secfluct}), since these are now being measured experimentally to very high precision. For smaller systems and high momenta, the predictive power of QCD was always in inter-momentum correlations, s(e.g., the ratio of $3jet/2jet$ events) rather than in the momentum spectrum alone.  We will see how Tsallis statistics will do there.
\subsection{Relation to the QCD phase diagram \label{secgamma}}
To proceed from acknowledging that the thermal model works to showing {\em what} it says about the chemical content of the system, it is likewise necessary to specify {\em which} thermal model to use.

\begin{figure}[t]
\epsfig{file=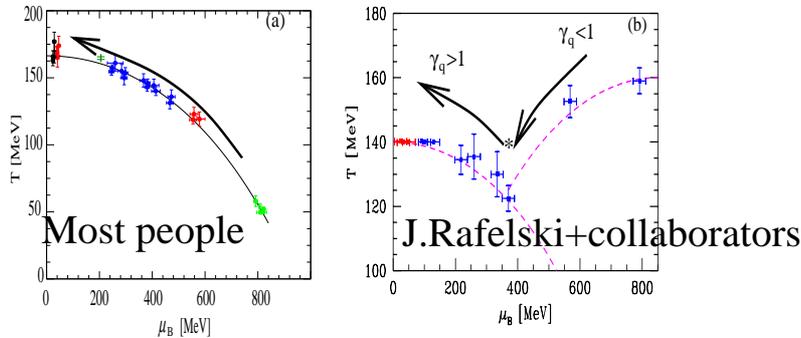,width=11cm}
\caption{\label{phaseall} ``Freeze-out'' phase diagram in equilibrium \cite{cleymans} and non-equilibrium \cite{petran} }
\end{figure}

The simplest is the one where chemical potentials are assigned to conserved charges {\em only}.   If this is done \cite{cleymans}, one obtains the phase diagram at the left hand side of Fig. \ref{phaseall}, with a decrease in baryochemical potential with increasing energy that fits a universal hadronization condition of $\ave{Energy}/\ave{particle} \sim 1$ GeV.
On the other hand, \cite{rafelski,petran} argue that one can not assume chemical equilibrium for hadrons when these are produced from hadronization of previously equilibrated quarks and gluons, because the strangeness and entropy content of the hadronic and QGP phases is very different.   One must therefore abandon detailed balance and assign also $\gamma_q \ne 1$ ($\gamma_q = \gamma_{\overline{q}}$) as chemical potentials, parametrizing the departure from detailed balance (where $\gamma=1$ and $\mu_i = \mu_{\overline{i}}$).
Both fits including and excluding $\gamma$ parameters give results  generally in agreement with the expectation of the fitters, with equilibrium fits showing a diagram that closely matches the phase diagram \cite{cleymans} while non-equilibrium shows a sharp jump in $\gamma$ \cite{petran,rafelski} possibly related to the onset of deconfinement \cite{gazdzicki,odyniec}.   The values of the bulk parameters of this system above the jump are then the expected values of the deconfined phase \cite{petran}.

Currently, the ultimate choice of models is based on  theoretical prejudice, as no experimental observable can distinguish between the two scenarios {\em conclusively}.  Particle ratios sensitive to $\gamma$ (resonances, nuclei, hypernuclei) generally are sensitive to post-chemical freezeout dynamics, which is expected to be longer in the equilibrium than in the non-equilibrium model, and whose real entity is not as yet ascertained.   

\begin{figure}[t]
\epsfig{file=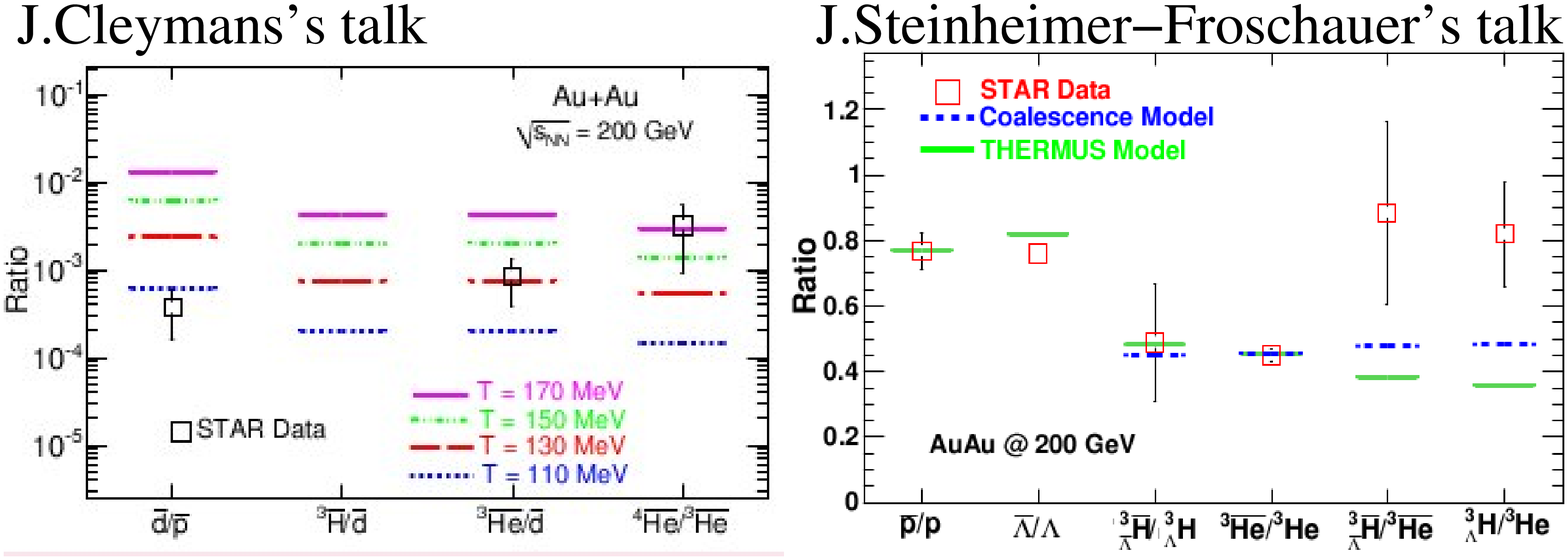,width=11cm}
\caption{\label{nuclei} Nuclei in heavy ion collisions compared to thermal model expectations}
\end{figure}

We shall see what the recent failure to fit $p/\pi$ in the equilibrium model \cite{prino} implies.   The statistical model analysis of nuclei, antinuclei and hypernuclei (Fig. \ref{nuclei}) \cite{cleymans,steinheimer} might be a very sensitive test, both for their sensitivity to $\gamma$ and their fragility in a hypothetical \cite{blume} hadronic reinteraction phase.
\section{Strangeness enhancement}
This topic, of course, is the historic origin of this meeting.  The idea, due to the authors of \cite{rafelski,mueller,koch} of using strangeness as a signature for deconfinement comes from observing that in ``a box of QGP'' close to $T_c$ strangeness will thermalize faster, {\em and equilibrium strangeness will be more dense}, than in a corresponding ``box of hadrons''.
Experimentally, then, the right variable is the ``enhancement'', strangeness abundance, normalized by the number of participants, in $A-A$ compared to a smaller system.    

\begin{figure}[t]
\begin{center}
\epsfig{width=10cm,figure=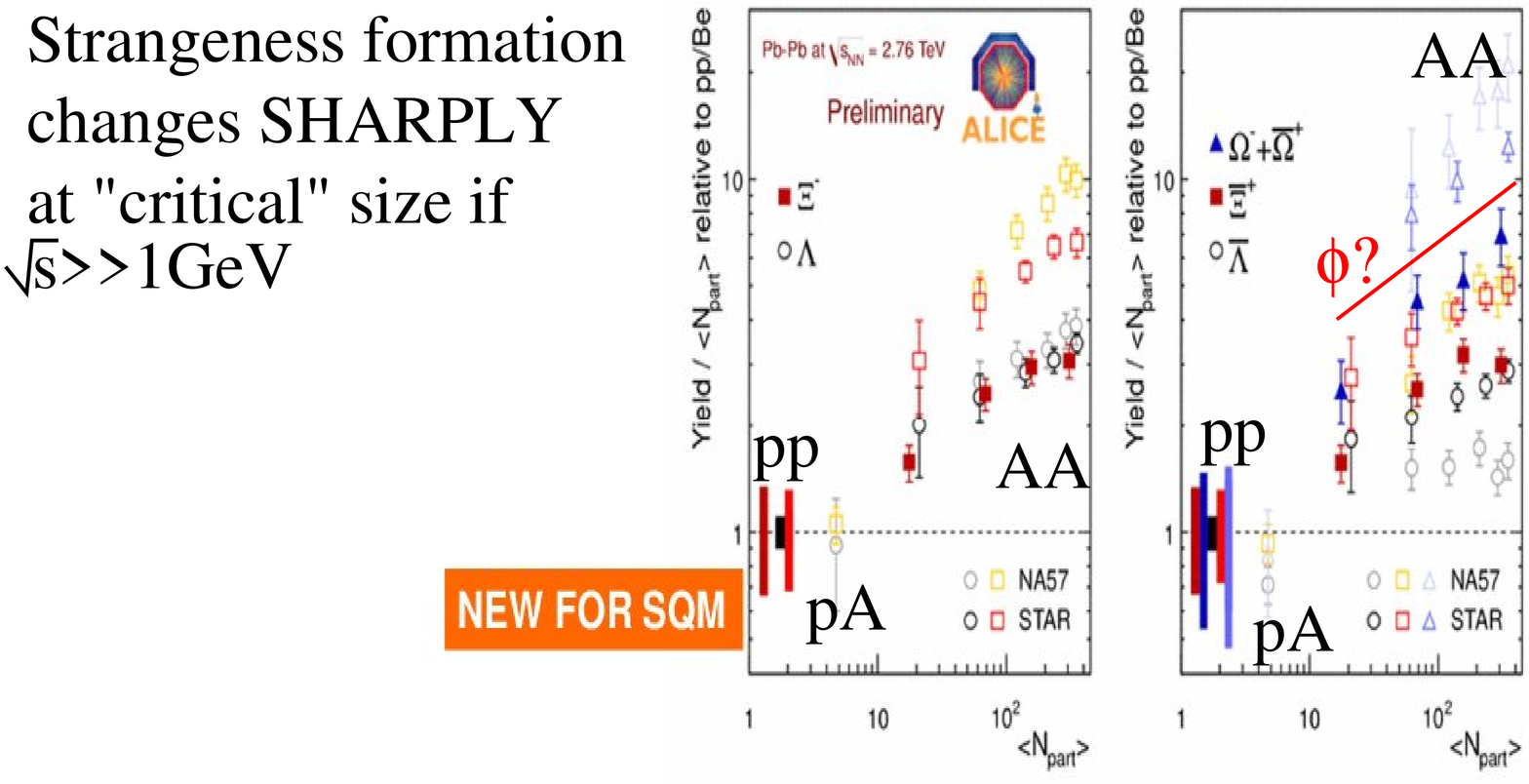}
\caption{\label{enhancement} Strangeness enhancement at the LHC and lower energies}
\end{center}
\end{figure}

Experimentally, there is no doubt:  Strangeness enhancement {\em is there!} \cite{prino} Fig. \ref{enhancement}.
It also seems clear, as previously discussed at SQM \cite{sqm07,torrieri}, that this enhancement is due to a change in ``intensive'' parameters (such as $\gamma_s$, defined in in the previous section \ref{secgamma}) rather than extensive ones (the ``canonical suppression'' in smaller systems described in \cite{oeschler}).   Further $\phi$ enhancement measurements are however needed to ensure this is true at {\em all} energies, particularly lower ones where energy-momentum conservation become more important.
At the LHC, the scaling difficulties mentioned in section \ref{secglobal} rear their head when interpreting the energy dependence of Fig. \ref{enhancement}.   Is enhancement at the LHC smaller due to some influence of canonical extensive variables, or is it because the {\em whole system} got larger by a power-law scaling, both in $p-p$ and $A-A$, but with quantitatively different powers?
The exciting culmination of this story, necessary to link Fig \ref{enhancement} to Fig. \ref{goodsig}, is at {\em low} energy \cite{odyniec,sorin,bleicher,gazdzicki}.   Does multistrange particle enhancement turn off at some low energy, as the deconfinement/$\gamma_s$ interpretation holds, or is it always there, as expected from domination of thresholds?   Since the Coulomb barrier (and energies of interest for deconfinement) are well separated from the threshold of producing the $\Omega$, investigating how multistrangeness enhancement behaves at lower energies will be crucial in ascertaining whether strangeness enhancement is indeed what is required for Fig. \ref{goodsig}.  
\section{Quarkonium suppression: The other smoking gun?}
\begin{figure}[t]
\epsfig{file=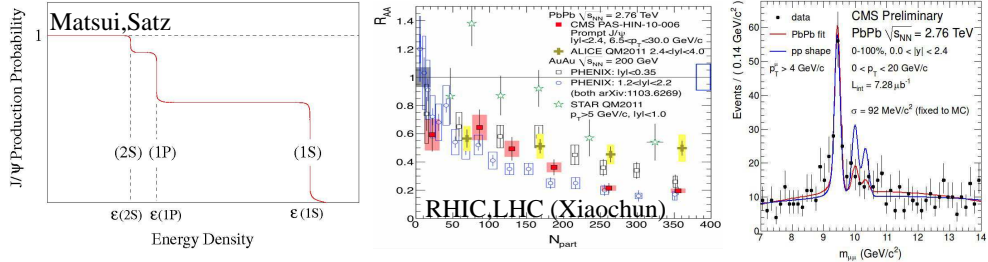,width=13cm}
\caption{\label{jpsiyes} Theory \cite{satz} (left panel) and experimental data \cite{xiaochun,hong} for quarkonium dissociation}
\end{figure}

\begin{figure}[t]
\epsfig{file=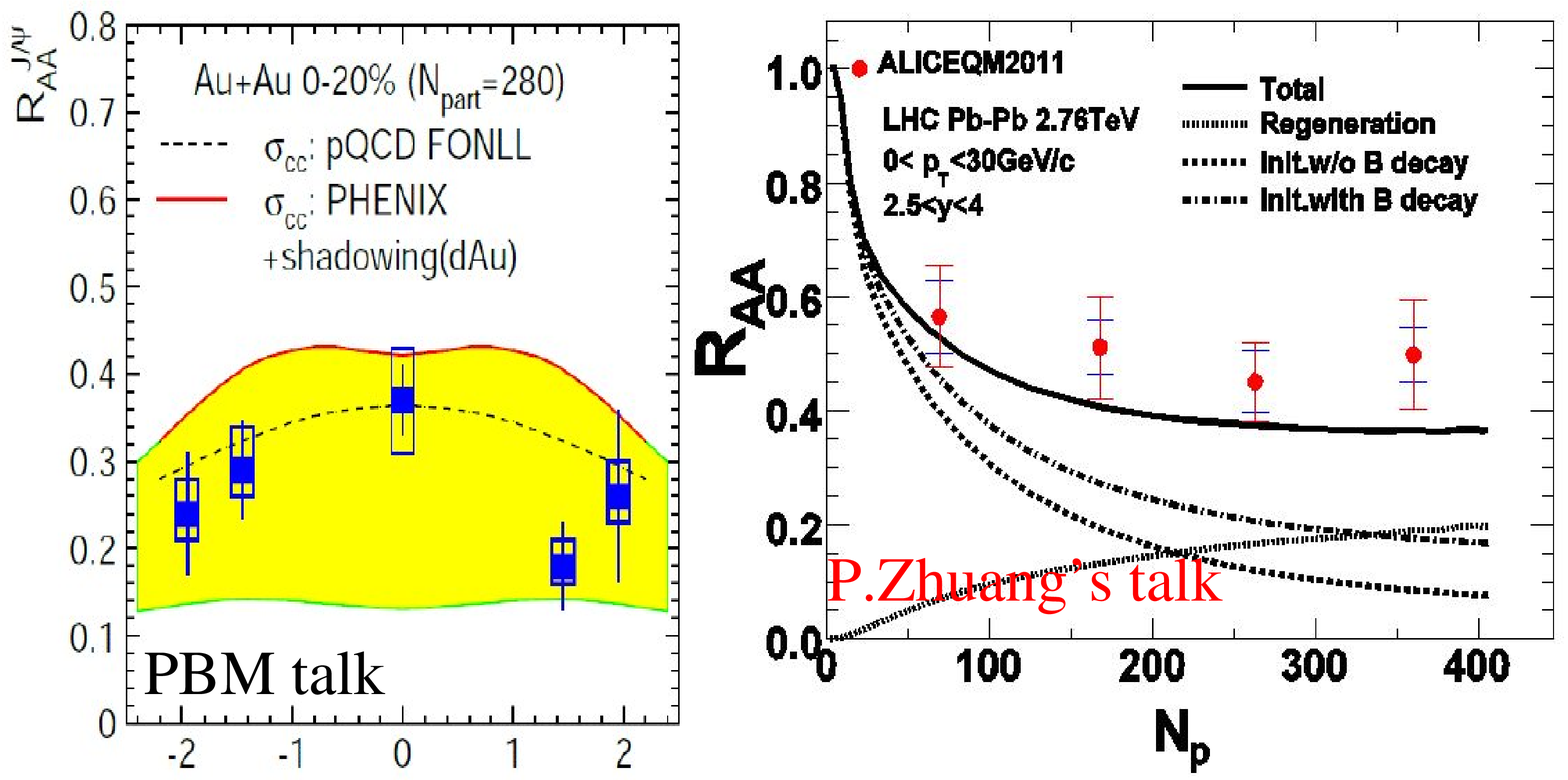,width=9cm}
\caption{\label{pengfei} Left panel: Quarkonium $R_{AA}$ as a function of rapidity \cite{pengfei} Right panel: Illustration of different effects going into quarkonium production \cite{pbm}}
\end{figure}

The other historical signature of deconfinement is Matsui and Satz's quarkonium suppression.  While quarkonium states are expected to survive deconfinement since the mass of their constituents $\gg \lqcd$, they will eventually break up due to Debye screening soon afterwards.   Indeed, some very beautiful experimental data has been presented here \cite{xiaochun,hong} (Fig. \ref{jpsiyes}) conclusively showing both charmonium suppression and even {\em sequential bottomonium dissociation } (excited $\Upsilon$ decay faster than the ground state).
However, as can be seen in the middle panel of Fig. \ref{jpsiyes}, quarkonium dissociation data's scaling is much more messy than expected if dissociation was the {\em only} factor at play \cite{pbm}.   That this is indeed the case is shown beyond reasonable doubt by the fact that $R_{AA}(y>0)<R_{AA} (y=0)$ \cite{pbm} (left panel of Fig. \ref{pengfei}, $R_{AA}$ is the suppression factor of hard probes), despite the fact that temperature in the fragmentation region can not be possibly higher than the temperature at midrapidity \cite{pengfei,pbm}.
This does not, by itself, invalidate the ``thermometer view'' of quarkonium.  It does, however, mean that thermal dissociation cannot be the {\em only} effect at play.   Initial production, dissociation, and hadronization (quarkonium coalescence \cite{pbm,rafelski}) play their part and need to be studied together \cite{pengfei}.

Due to this, quarkonium as a deconfinement smoking-gun a la Fig. \ref{goodsig} seems doubtful.  Thermal model applicability \cite{rafelski,pbm}, however, means reality could be simpler than transport-based models such as \cite{pengfei,lang,bratkovskaya} imply.  
\section{Fluctuations, higher cumulants, and their scaling \label{secfluct}}
By definition, \underline{statistical} mechanics of any kind assumes cumulants of an event-by-event observable scale in a computable way.
Thus, any statistical behavior should, in principle, be constrained by measuring higher cumulants.   In particular, a strong peak in fluctuations and higher cumulants would be a conclusive evidence for critical point behavior \cite{redlich,ritter,gazdzicki}.

In the absence of such a peak, one can use experimentally measured fluctuations to see if statistical models can describe more than particle averages.
In practice, however, measurements beyond averages are beset by difficulties specific to heavy ion collisions (effect from hadronizing jets, detector acceptance, cumulants in the distribution of event size, and so on).   In general, for fluctuations, such difficulties never go away but are minimized by choosing a good fluctuation observable.   $\sigma_{dyn}$ and $\nu_{dyn}$ of particle ratios would be good candidates \cite{torrieri,kasia,tarnowsky}.

\begin{figure}[t]
\epsfig{file=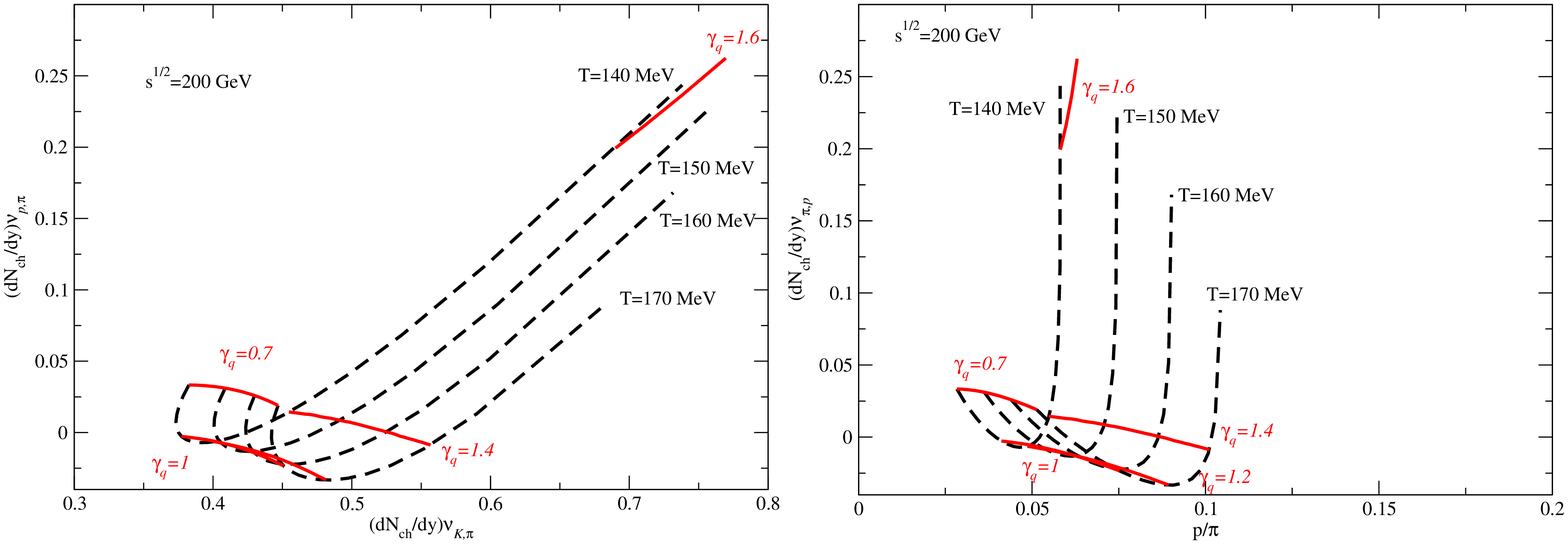,width=13cm}
\caption{\label{figmenew} Sensitivity of different fluctuation observables to different parameters, using $T,\mu$ from fits at 200 GeV}
\end{figure}

Preliminary results \cite{tarnowsky} suggest that fluctuations are lower than all thermal models, with equilibrium being more (but not totally) compatible with data than non equilibrium.  As these preliminary results are in marked contrast from previous published results of the same collaboration, this is not as yet a definite conclusion.   It should be noted that, due to BE corrections \cite{torrieri}, pion fluctuations (and hence fluctuations of ratios involving pions) are {\em very} sensitive to the exact value of $\gamma_q$. 10$\%$ variations of $\gamma_q$, well below the error in fits of \cite{petran,rafelski}, could easily increase and decrease fluctuations by $\sim 100\%$ (Fig. \ref{figmenew}).
Thus, the finish line in this story is visible, but we are still moving towards it, and {\em it} might be moving too!
Regarding the more ``spectacular'' use of fluctuations, for critical point searching, there are further theoretical difficulties, well illustrated in \cite{herold,bleicher}.

\begin{figure}[t]
\begin{center}
\epsfig{width=17cm,figure=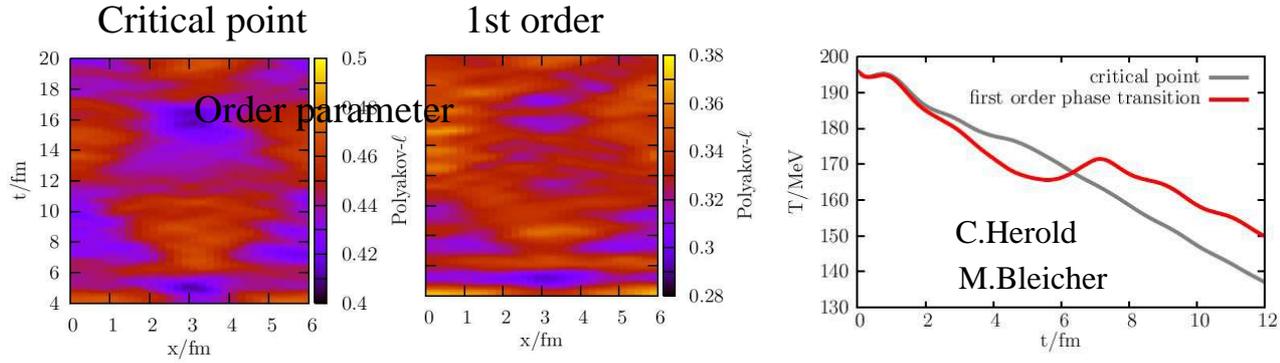}
\caption{\label{herold} Hydrodynamics in the vicinity of a critical point taking fluctuation and dissipation into account \cite{herold}}
\end{center}
\end{figure}

A critical theoretic uncertainity is the interplay between the ``diverging'' order parameter fluctuations, relaxation time (controlling both dissipation and stochastic noise), bulk evolution, and hydrodynamic fluctuations (see also \cite{kapusta}).
In addition, bubble formation beyond the critical point, in the first order region of the phase diagram, might well enhance fluctuations on its own.
The model developed by \cite{herold} and collaborators certainly impressively handles most of these effects (Fig. \ref{herold}), but a lot of work needs to be done to conclusively answer the question of whether the critical point will be seen by a large fluctuations enhancement.

One way to increase the expected signal of the divergence is to use higher cumulants rather than fluctuations (a reapplication of the ``factorial moments'' idea , which, being in Krakow \cite{bialas}, we should be familiar with!).  On the lattice, these higher cumulants do indeed seem to behave as if a critical point might be approaching \cite{redlich,gavai,schmidt}, although, once again, the evidence is {\em not} conclusive of its existence.  However, existing measurements have  allowed \cite{ritter} a comparison between lattice QCD and experimental data.

While the results look successful, estimating the theoretical systematic errors of the results in \cite{ritter} is difficult because a higher cumulant measurement is a measurement of {\em unusual events}:  The effect of 
\begin{itemize}
\item acceptance and particle misidentification
\item jet fragmentations to low momentum particles
\item conservation laws ( $P_{B-\overline{B}>A}=0$ in experiment, a Grand canonical tail stratching to $B-\overline{B}=\infty$ on the lattice)
\item cumulants of the freeze-out volume distribution (whatever that is!)
\end{itemize}
is potentially enough to completely overturn such comparisons.   For fluctuations, the effect of these can be partially mitigated by choosing the right fluctuation observable (e.g. $\sigma_{dyn}$) and doing mixed event corrections (and even this is controversial \cite{tarnowsky,kasia,torrieri}), but such issues have not as yet been explored for higher cumulants.
\section{$v_2$: From perfect liquid to phase transition?}
Elliptic flow at RHIC has certainly been a huge success story: The discovery of the perfect liquid aroused enormous theoretical interest, and opened the door to comparisons between our field and other experimental (cold atoms,unstable plasmas \cite{deja}) and theoretical (strongly coupled N=4 SYM plasma) good fluids.   Still missing is a robust connection between fluid behavior and deconfinement, preferably via a Fig. \ref{goodsig} for fluids.   To do this, one would have to see how $v_2$ turns on and off as $\sqrt{s}$ is lowered.

\begin{figure}[t]
\begin{center}
\epsfig{width=16cm,figure=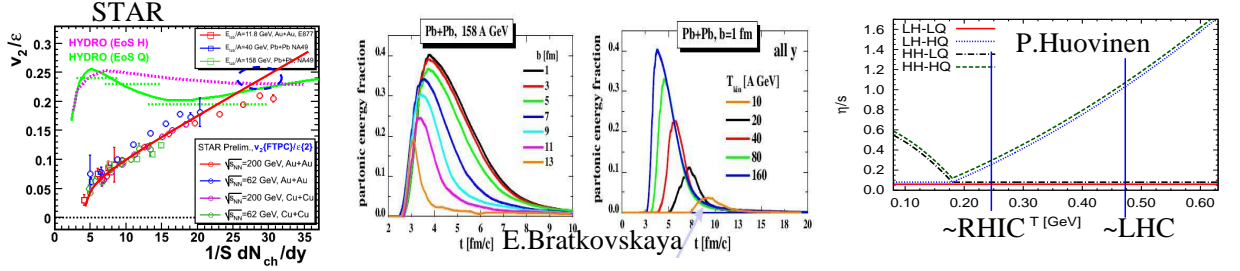}
\caption{\label{v2fig} $v_2$ scaling with multiplicity (left panel \cite{odyniec}) and partonic to hadronic contributions as a function of energy in transport (middle panel \cite{bratkovskaya}) and a viscosity parametrization used in hydro (right panel \cite{huovinen})}
\end{center}
\end{figure}

Data in this direction is finally starting to appear.  As shown by \cite{odyniec}, at $\sqrts=39$ GeV, quark coalescence behavior of $v_2$ starts to definitely break, as particle/antiparticle $v_2$ difference grows.  On the other hand, quark number scaling also seems to be broken at the LHC.
More generally, the considerations made in section \ref{secglobal} prevent any straight-forward link between $v_2$ scaling and deconfinement, as transparency and jet production and absorption {\em certainly} give an imprint to $v_2$ and vary strongly with energy.  In fact, it is puzzling how {\em well} the {\em absolute} $v_2$ scales (Fig. \ref{v2fig} left panel \cite{odyniec}):

From a theoretical side, \cite{huovinen,bratkovskaya} have shown that between our low energies ($\sqrts=7$ GeV) and LHC energies, the partonic contribution of the initial state varies from negligible to dominant.   It is also highly likely that transport properties of partons and hadrons are very different (they are in both \cite{huovinen} and \cite{bratkovskaya}).    It remains to be seen whether the good experimental scaling (Fig. \ref{v2fig} left panel) is compatible with such variation (Fig. \ref{v2fig} right panel) in the intensive properties of the system at different energies/centralities. 
\section{``Less promising'' observables}
By this name, I mean less promising from the point of view of Fig. \ref{goodsig}.  It is not clear what, if anything, do these observables have to do with a phase transition and a change in degrees of freedom.   The very interesting results presented here for these observables, however, could change the situation.
\subsection{Momentum correlations (ridges, cones, and $v_n$)}
It has become justifiably fashionable to describe 2-particle correlations in terms of higher Fourier components of the particle distribution w.r.t. reaction plane.   Indeed, the ``ridge'' and the ``away-side peak'' both at RHIC and the LHC can be very well decomposed into Fourier harmonics, of which the most important is $v_3$ \cite{prino,kabana,wosiek}.   To go from there to claiming all such correlations are generated hydrodynamically, via higher moments, is a possibility but not a certainty.  As remarked in \cite{wang}, ``$v_n$s are a Fourier transform, not a theory''

\begin{figure}[t]
\begin{center}
\epsfig{width=12cm,figure=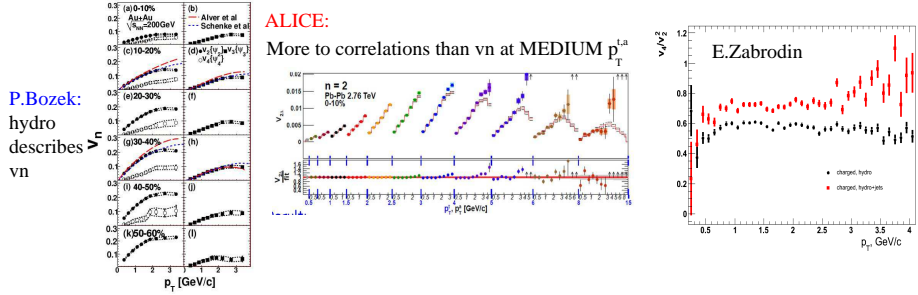}
\caption{\label{vn} Left panel:  Hydrodynamics seems to describe all $n_n$ coefficients \cite{bozek}.  Center panel 2-particle correlations are however not simply given by a sum of $vn$s \cite{prino}.  Right panel: Nor should they be \cite{zabrodin} }
\end{center}
\end{figure}

While $v_n$s are expected to be produced in hydrodynamics due to fluctuating initial conditions \cite{bozek}, the effect of jet fragmentation (also correlated with reaction planes due to jet energy loss) and jet-medium interactions (Mach cones?) can not be excluded either.
Disentangling these is not a trivial matter \cite{zabrodin}, but it can already be seen \cite{prino} that $v_n$s are {\em can not} be a function of reaction plane {\em alone}, and non-reaction plane jetty contributions rear their head when the associated and trigger momenta are $\sim 2$ GeV, at the higher end of hydrodynamics (Left and middle panel of Fig. \ref{vn}).
Saying something further which is both {\em certain} and {\em model-independent} is difficult, as \cite{zabrodin} (right panel of Fig. \ref{vn}) has shown.

Perhaps more discerning observables might help clarify the situation.
Correlations between a {\em heavy quark trigger} and a {\em soft light quark} can not be due to hydrodynamic correlations alone, since a hotspot {\em can} produce higher momentum particles but {\em can not} produce a heavy quark unless a ``hard'' process is also involved (hence, the heavy quark direction is not determined by the flow, and the heavy quark remains a ``jet'', a former high energy parton even if it loses all momentum).

Alternatively, correlating the shape of the correlation function to the global characteristics of the event is also promising:  If the ridge is generated by hotspot-type correlations focused by transverse flow (as in \cite{bozek}), events with a higher $\left. \ave{ p_T} \right|_{event}$ will have narrower (more focused) ridges.  This is not true, or at least not necessarily true, if the ridge is due to local dynamics.  Studies in this direction are progressing \cite{chanaka}.
\subsection{Jet suppression}
The field of jet suppression has been very active on a theoretical level, but as yet inconclusive in comparing to data.   Very briefly, as can be seen in \cite{prino,kabana,otwinowski}, pQCD works comparatively well for $R_{AA}^{light}$ but tends to miss $v_2$ of these quarks for the same values of the opacity parameter.
$R_{AA}^{heavy}$ is generally not well described with the same parameters, unless transport parameters assume unrealistically (for the applicability of pQCD) high values.

This has given rise to strongly coupled approaches, both in terms of AdS/CFT (which, in its on-shell limit, does a better job for $v_2$ \cite{otwinowski}), off-shell transport \cite{bratkovskaya,aichelin}, and many-body effects \cite{uphoff}.

Thus, at the moment we do not know if the system is weakly or strongly coupled,
and whether the strong coupling admits a quasiparticle description or not.  As to the relation between opacity to jets and deconfinement, this has not as yet been theoretically investigated.

\begin{figure}[t]
\epsfig{file=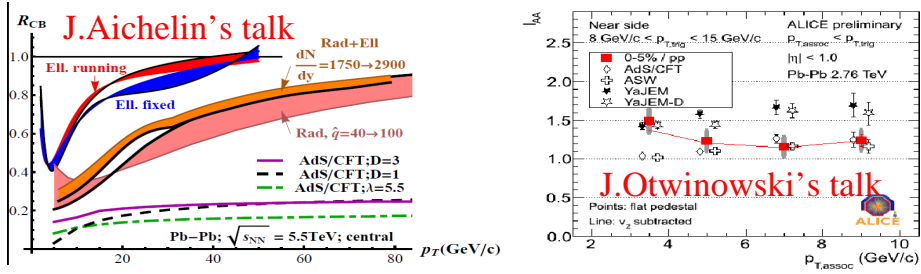,width=12cm}
\caption{\label{jets} $R_{AA}$ of charm and bottom jets in different models of jet energy loss }
\end{figure}

It is clear that, to go towards falsification, some effort is needed to construct smart observables.  The scaling of $R_{AA}$ with quark mass, as shown in Fig. \ref{jets} \cite{aichelin}, or distinguishing quark from gluon jets \cite{pochybova} are promising avenues.
Whether, and to what extent this can be done at lower energies, to explore the relationship between jet energy loss and deconfinement, is still an open experimental question.
\section{Conclusions} 
The only conclusion that can be made is that this meeting has been extremely interesting and productive.  Fig. \ref{goodsig} is not there yet, but I see no reason for {\em excluding} its existence.   And if its there, we are on the way to find it.   
Thank you to the organizers for the smooth running of this conference, and a very happy jubilee to Jan Rafelski, without whom these meetings would probably not place.   To be continued, Birmingham, 2013 \cite{villalobos}!

G.~T.~acknowledges the financial support received from the Helmholtz International
Centre for FAIR within the framework of the LOEWE program
(Landesoffensive zur Entwicklung Wissenschaftlich-\"Okonomischer
Exzellenz) launched by the State of Hesse, the organizers of SQM for their generous support, and Johann Rafelski for the years of collaboration, discussion, and support given.

\end{document}